\documentclass[12pt]{iopart}        

\usepackage{iopams}
\usepackage{graphicx}

\begin{document}

\title[Thermal expansion of
$RT${\rm Mg} ($R = {\rm Eu}$, {\rm Gd} and $T = {\rm Ag}$, {\rm
Au})]{Thermal expansion of the magnetically ordering
intermetallics $RT$Mg ($R = $ Eu, Gd and $T = $ Ag, Au)}

\author{J~Rohrkamp$^1$, O~Heyer$^1$, T~Fickenscher$^2$, R~P\"{o}ttgen$^2$,
S~Jodlauk$^{1,3}$, H~Hartmann$^{1,4}$, T~Lorenz$^{1,5}$ and
J~A~Mydosh$^{1}$\footnote[0]{\noindent $^3$Present address:
Institut f\"{u}r Kristallographie, Universit\"{a}t zu K\"{o}ln, Z\"{u}lpicher
Str.\ 49b, 50674 K\"{o}ln.\\ $^4$Present address: DLR K\"{o}ln, Institut
f\"{u}r Materialphysik im
Weltraum, Linder H\"{o}he, 51147 K\"{o}ln. \\
$^5$ Author to whom correspondence should be addressed
(lorenz@ph2.uni-koeln.de).} }

\address{$^1$ II.\,Physikalisches Institut, Universit\"{a}t zu K\"{o}ln, Z\"{u}lpicher
Str.\ 77, 50937 K\"{o}ln, Germany}

\address{$^2$ Institut f\"{u}r Anorganische und Analytische Chemie,
Westf\"{a}lische Wilhelms-Universit\"{a}t M\"{u}nster, Correnstr.\ 30, 48149
M\"{u}nster, Germany}

\begin{abstract}
We report measurements of the thermal expansion for two Eu$^{+2}$-
and two Gd$^{+3}$-based intermetallics which exhibit ferro- or
antiferromagnetic phase transitions. These materials show sharp
positive (EuAgMg and GdAuMg) and negative (EuAuMg and GdAgMg)
peaks in the temperature dependence of the thermal expansion
coefficient $\alpha$ which become smeared and/or displaced in an
external magnetic field. Together with specific heat data we
determine the initial pressure dependences  of the transition
temperatures at ambient pressure using the Ehrenfest or
Clausius-Clapeyron relation. We find large pressure dependences
indicating strong spin-phonon coupling, in particular for GdAgMg
and EuAuMg where a quantum phase transition might be reached at
moderate pressures of a few~GPa.
\end{abstract}





There has been a continuing interest in the equiatomic ternary
intermetallic compounds based upon Eu and Gd rare earths, the
noble metals Ag and Au, and Mg.\cite{P1} In particular through the
past ten years their synthesis, chemical and physical properties
have been studied in some detail.\cite{P2,P3,P4,P5,P6,P7,P8,P9}
The Eu-based crystallize with the orthorhombic TiNiSi structure,
while the Gd-based materials adopt the ZrNiAl structure. Recently,
the electronic structure has been determined by x-ray
photoelectron spectroscopy and compared with LDA+U band structure
calculations.\cite{P8} Here the valency, Eu$^{+2}$ and Gd$^{+3}$,
was firmly established and a variety of $s,p,d$-conduction bands
were found at the Fermi level. One unusual feature of the
photoemission spectra was the localized nature of the Ag and Au
$d$-bands below $E_{\rm F}$. Since both Eu$^{+2}$ and Gd$^{+3}$
possess large magnetic moments ($S = 7/2$), their magnetic
ordering properties, mediated by the oscillating amplitude RKKY
interaction, may strongly depend on spin-orbit and spin-lattice
couplings. A recent investigation has characterized many of the
bulk thermodynamic and transport properties related to the
ferromagnetic transitions of EuAgMg, EuAuMg and GdAgMg and the
antiferromagnetic one of GdAuMg.\cite{P9} Information about the
above-mentioned magnetoelastic behavior can be obtained from
measurements of the thermal expansion.

In this work we present the results of thermal expansion measured
over a large temperature range (4 to 200~K) in applied magnetic
fields reaching 14~T. We observe sharp peaks at the magnetic phase
transitions in the uniaxial thermal expansion $\alpha(T) =
\frac{1}{L_0} \frac{\partial \Delta L}{\partial T}$, where $L_0$
is the sample length and $\Delta L$ its temperature-induced
change. In all four compounds we observe pronounced anomalies
$\Delta \alpha$ at the magnetic ordering transitions. For EuAgMg
and GdAuMg the sign of $\Delta \alpha$ is positive, while it is
negative for EuAuMg and GdAgMg. These anomalies are smeared and/or
displaced upon applying the magnetic field. By comparing $\Delta
\alpha$ with the corresponding anomalies $\Delta C_p$ of the
specific heat~\cite{P9} at $T_{\rm C,N}$ we obtain the pressure
dependences of the critical temperatures for a first- or
second-order phase transition via the Clausius-Clapeyron or the
Ehrenfest relation, respectively. The results are most
interesting for the ferromagnets GdAgMg and EuAuMg, where we find
a first- and a second-order phase transition, respectively, with
very large negative pressure dependences of $T_{\rm C}$. In both
materials, a finite pressure of the order of a few~GPa should
drive $T_{\rm C}$ to zero suggesting pressure-induced quantum
phase transitions.

Polycrystalline samples of the above compounds were synthesized,
annealed and characterized as described
previously.\cite{P7,P8,P9} The thermal expansion was measured in a
capacitance dilatometer inserted in a $^4$He gas-flow cryostat
covering a temperature range from about 2 to 300~K.\cite{P10} By
using a superconducting magnet, fields up to 14~T could be
applied over the entire temperature range. Accordingly, the
length changes $\Delta L/L_0$ could be accurately detected
through the magnetic transitions with and without the external
field.

\begin{figure}[t]
\hfill
\includegraphics[width=0.85\textwidth]{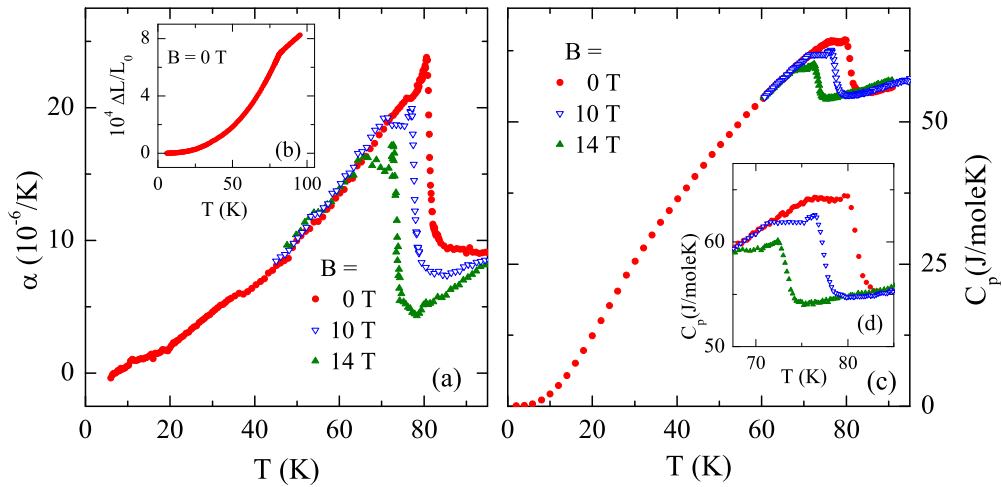}
\caption{\label{gdau} (a) Thermal expansion $\alpha$, (b) relative
length change $\Delta L/L_0$, and (c,d) specific heat of
antiferromagnetic GdAuMg for different magnetic fields.}
\end{figure}

Figure~\ref{gdau}(a) displays $\alpha(T)$ in various magnetic
fields for GdAuMg. The antiferromagnetic ordering at $T_{\rm N
}=81$~K causes a step-like anomaly $\Delta\alpha(T)$ of positive
sign, which is typical for a second-order phase transition
without pronounced fluctuations. A very similar anomaly is
present in the specific heat as shown in figure~\ref{gdau}(c). The
enhanced thermal expansion below $T_{\rm N}$ signals a
spontaneous contraction in the magnetically ordered phase, which
is also directly seen in $\Delta L(T)/L_0$ displayed in
figure~\ref{gdau}(c). The application of a magnetic field causes
a systematic shift of the anomalies $\Delta \alpha$ and $\Delta
C_p$ towards lower temperature, signaling a decreasing $T_{\rm
N}$. As expected the antiferromagnetic order is destabilized by a
magnetic field, although the rate $\partial T_{\rm N}/\partial B
\simeq -0.4$~K/T in 14~T appears relatively large in view of the
high $T_{\rm N}$.

\begin{figure}[t]
\hfill
\includegraphics[width=0.85\textwidth]{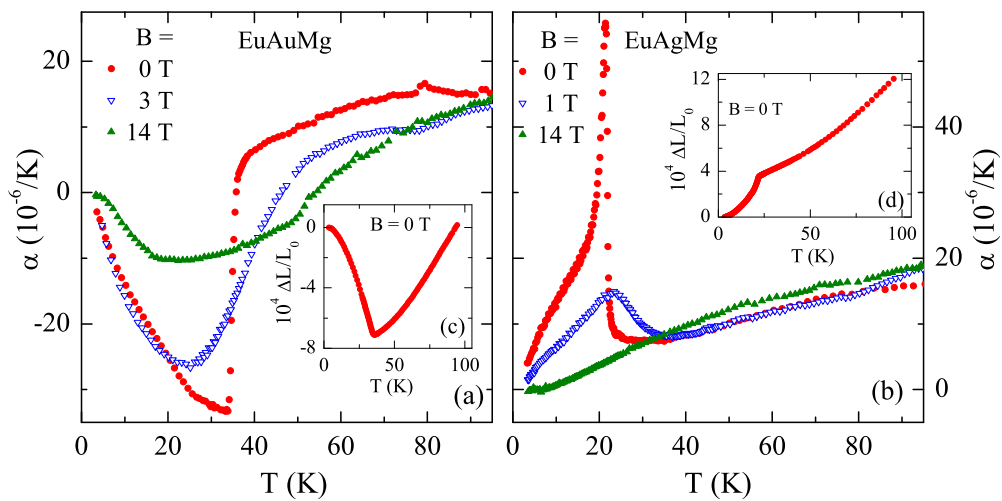}
\caption{\label{eu} Thermal expansion $\alpha$ of the
ferromagnetic compounds (a)~EuAuMg and (b)~EuAgMg for various
magnetic fields. The respective insets~(c) and~(d) show the
relative length changes $\Delta L/L_0$ for zero field.}
\end{figure}

Figure~\ref{eu} shows $\alpha(T)$ in various fields for EuAuMg and
EuAgMg, which order ferromagnetically at $T_{\rm C}=35$~K and
22~K, respectively. In zero field we again find a step-like
anomaly $\Delta \alpha$ for EuAuMg. However, in contrast to
GdAuMg the sign of $\Delta \alpha$ is negative meaning that the
ferromagnetic ordering in EuAuMg is accompanied by a spontaneous
expansion as shown in figure~\ref{eu}(c). For EuAgMg the sign of
the zero-field anomaly $\Delta \alpha$ is again positive, i.e.\ a
spontaneous contraction occurs, but in this case $\Delta \alpha$
has a lambda-like shape as it is typical for a second-order phase
transition where the fluctuations are more pronounced. Such a
difference in the anomaly shapes is also present in the specific
heat anomalies.\cite{P9} For both Eu-based compounds the
application of a magnetic field causes a drastic broadening of the
anomalies. This is typical for a ferromagnet because a sizeable
dmagnetic field induces a strong magnetization already well above
the zero-field $T_{\rm C}$. As a consequence the magnetization
does not develop spontaneously below a critical temperature
anymore and, strictly speaking, a ferromagnetic transition
temperature can only be defined for zero magnetic field.

\begin{figure}[t]
\hfill
\includegraphics[width=0.85\textwidth]{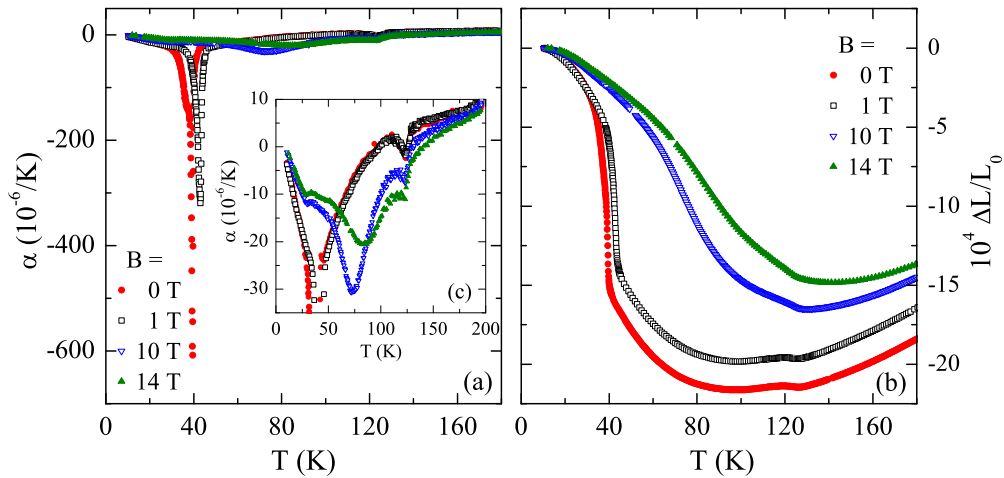}
\caption{\label{gdag} (a) Thermal expansion $\alpha$ and (b)
relative length changes $\Delta L/L_0$~(b) of the ferromagnet
GdAgMg for various magnetic fields. The inset~(c) shows an
expanded view of $\alpha$ in order to visualize the additional
anomaly at 125~K.}
\end{figure}

In the ferromagnet GdAgMg we observe a huge anomaly
$\Delta\alpha$ of negative sign at $T_{\rm C}=39.5$~K as is
illustrated in figure~\ref{gdag}(a). The shape of the $\alpha$
anomaly corresponds to an almost jump-like change of $\Delta
L/L_0$ as shown in figure~\ref{gdag}(b). This is a clear
indication for a first-order transition in GdAgMg, in agreement
with our conclusions from the specific heat measurements, where a
very similar anomaly shape is present at $T_{\rm C}$.\cite{P9}
Moreover, we observe an additional smaller anomaly of $\alpha$ at
$T^{\star}\simeq 125$~K which hardly changes in an applied
magnetic field, see figure~\ref{gdag}(c). In contrast, the sharp
low-temperature anomaly drastically broadens and shifts upwards
in temperature. In the highest field this broadening even extends
to temperatures above about 175~K, i.e.\ to $T\gg T^{\star}$.
Based on an analysis of the entropy change we suspected that the
complete magnetic ordering in GdAgMg might be achieved via both
transitions at $T_{\rm C}$ and $T^{\star}$.\cite{P9} However, the
transition at $T^{\star}$ is not seen in our measurements of the
magnetic susceptibility $\chi$. Above about 50~K, $\chi$ shows
clear Curie-Weiss behavior, i.e.\ $\chi(T)=\mu_{eff}^2/3k_{\rm
B}(T-\Theta)$ with $\Theta=39.7$~K and $\mu_{eff}=7.96\mu_{\rm
B}$ close to the expected values of a ferromagnet with an $S=7/2$
and $T_{\rm C}=39.5$~K. This and the very different field
dependences, in particular the fact that the field-induced
broadening of the $T_{\rm C}$ anomaly even exceeds to $T\gg
T^{\star}$, makes a common origin of both anomalies very unlikely.

Since $T^{\star}$ does not change with field, one may suspect a
structural origin of this anomaly, however, no low-temperature
diffraction data have been recorded yet. Alternatively, the
$T^{\star}$ anomaly could arise from small amounts of impurity
phases, e.g.\ GdAg and GdMg. Although our sample shows phase
purity in x-ray powder diffraction, we cannot exclude the
presence of a few~\% of such impurity phases. Both of these
binary intermetallic compounds exhibit magnetic phase transitions
between 96 and 130~K.\cite{P11,P12} In GdMg a ferromagnetic order
occurs at 120~K followed by a canting transition at 96~K, while a
single antiferromagnetic phase develops in GdAg below 133~K. The
presence of GdMg appears also to be unlikely, since even for a
few~\% of such an impurity phase the spontaneous magnetization
due to the ferromagnetic order should be visible in $\chi(T)$ and
the anomaly should broaden in a magnetic field. In contrast, the
presence of a few~\% of GdAg cannot be excluded from our data:
(i) due to the large background of the paramagnetic majority
phase the expected weak magnetization change at $T_{\rm N}$
cannot be resolved in $\chi(T)$, and (ii) magnetic fields up to
14T will not affect the antiferromagnetic order since the
ordering temperature is so high. Because the $T^{\star}$ anomaly
most probably arises from an impurity phase, it will not be
considered in the following discussion of the pressure
dependences.

The measurements of thermal expansion and specific heat at ambient
pressure allow to derive the initial slope of the change of
$T_{\rm N,C}$ under finite pressure using either the
Clausius-Clapeyron for a first-order transition
\begin{equation}
 \label{claus}
 \left.\frac{\partial T_{\rm N,C}}{\partial p}\right|_{p_0} = \frac{\Delta V}{\Delta
 S}\,\, ,
\end{equation}
or the Ehrenfest relation for a second-order phase transition.
\begin{equation}
 \left.\frac{\partial T_{\rm N,C}}{\partial p}\right|_{p_0} =
 3\,T_{\rm N,C}V_{mol}\frac{\Delta \alpha}{\Delta C_p}\,\, .
 \label{ehre}
\end{equation}
The jump-like changes of the volume and entropy are obtained via
integration $\Delta V=V_{mol}\,\int 3\,\alpha(T)\,{\rm d}T$ and
$\Delta S=V_{mol}\,\int C_p(T)/T \,{\rm d}T$, respectively. The
temperature range of integration around $T_{\rm N,C}$ reflects
the broadening of the first-order phase transition, which is
always finite in a real solid. We note that one has to use
$3\,\alpha $ here and in equation~(\ref{ehre}), since the
hydrostatic pressure dependence is related to the volume
expansion $\beta$, which is given by $\beta=3\,\alpha$ for a
homogeneous polycrystal.

\begin{table}[t]
 \caption{Some characteristic properties of $RT$Mg.
 \label{tab}}
 \vskip5mm
 \begin{center}
  \begin{tabular}{ccccc}
  \hline
    & $V_{\rm mol}$   & magnetic  & $T_{\rm N,C}$ & $\partial T_{\rm N,C}/\partial p$ \\
  material   & $\rm \left(cm^3/mole\right)$& order & $\rm \left(K\right)$ & $\rm \left(K/GPa\right)$ \\
 \hline
 GdAuMg  & 41.03 & antiferro & 81.0 & 12 \\
 GdAgMg  & 43.04 & ferro     & 39.5 & $-35$ \\
 EuAuMg  & 45.01 & ferro     & 35.0 & $-14$ \\
 EuAgMg  & 48.70 & ferro     & 22.0 & 9 \\
   \hline
 \end{tabular}
  \end{center}
  \end{table}

The obtained pressure dependences $\partial T_{\rm N,C}/\partial
p$ together with some other characteristic properties are given in
table~\ref{tab} for all four compounds. Obviously, we obtain
rather large absolute values of $\partial T_{\rm N,C}/\partial p$
of different signs depending on the sign of the respective
$\alpha$ anomaly. These large values confirm the presence of a
large spin-lattice coupling in these $RT$Mg compounds, which we
attribute to the oscillatory nature of the RKKY interaction. This
may also be the main reason for the variation from ferro- to
antiferromagnetic order and the wide range of different
transition temperatures in the different $RT$Mg compounds,
although their structures are not too different. We find the
largest pressure dependence in GdAgMg and associate this with the
first-order nature of the magnetic transition. In order to drive
a magnetic phase transition to first order a strong spin-lattice
coupling is required and this would also provide the mechanism
for a large pressure dependence $\partial T_{\rm N,C}/\partial p$.

We emphasize that our estimates of $\partial T_{\rm N,C}/\partial
p$ based on the Ehrenfest or the Clausius-Clapeyron relation only
yield the initial slopes at ambient pressure and extrapolations
of our data to finite pressure have to be treated with caution.
Nevertheless, the obtained negative values of $\partial T_{\rm
C}/\partial p =-14$ and $-35$~K/GPa for EuAuMg and GdAgMg,
respectively, are so large, that a complete suppression of the
ferromagnetic order may be reached for both compounds when a
rather moderate hydrostatic pressure of a few GPa is applied.
Therefore both compounds are interesting candidates where
pressure-induced quantum phase transitions can be studied.

In summary, due to the significant spin-lattice coupling we were
able to probe the magnetic ordering transitions of EuAgMg, GdAuMg,
EuAuMg and GdAgMg via thermal expansion measurements. Using also
specific heat data we were able to derive the initial pressure
shifts of $T_{\rm C,N}$, which are rather large for all four
compounds. The largest effects are observed in EuAuMg and GdAgMg
and in both materials the ferromagnetic order is expected to be
completely suppressed by hydrostatic pressure of a few GPa. Direct
high-pressure experiments of $T_{\rm N,C}$ on these compounds are
highly desirable.

\ack

We thank the
Degussa-H\"{u}ls AG for a generous gift of noble metals. This work
was supported by the Deutsche Forschungsgemeinschaft through the
priority programme SPP 1166 "Lanthanoidspezifische
Funktionalit\"{a}ten in Molek\"{u}l und Material".


\section*{References}


\begin{thebibliography}{99}

\bibitem{P1} Iandelli A 1994 {\em J. Alloys Compd.} {\bf 203} 137

\bibitem{P2} Fickenscher Th and P\"{o}ttgen R 2001 {\em J. Solid State Chem.} {\bf 161} 67

\bibitem{P3} Gibson B J, Das A, Kremer R K, Hoffmann R-D and
P\"{o}ttgen R 2002 {\em J. Phys.: Condens. Matter} {\bf 14} 5173

\bibitem{P4}  Johrendt D, Kotzyba G, Trill H, Mosel B D, Eckert H,
Fickenscher Th and P\"{o}ttgen R 2002 {\em J. Solid State Chem.} {\bf
164} 201

\bibitem{P5}  Latka K, Kmie\'{c} R, Pacyna A W, Tomkowicz T, Mishra R,
Fickenscher T, Piotrowski H, Hoffmann R-D and P\"{o}ttgen R 2002 {\em
J. Solid State Chem.} {\bf 168} 331

\bibitem{P6}  Kraft R, Fickenscher Th, Kotzyba G, Hoffmann R-D and
P\"{o}ttgen R 2003 {\em Intermetallics} {\bf 11} 111

\bibitem{P7} Latka K, Kmie\'{c} R, Pacyna A W, Fickenscher  Th,
Hoffmann  R-D and P\"{o}ttgen R 2004 {\em Solid State Sci.} {\bf 6}
301

\bibitem{P8}Gegner J, Koethe T C, Wu Hua, Hartmann H, Lorenz T,
Fickenscher T, P\"{o}ttgen R and Tjeng L H 2006 {\em Phys. Rev. B}
{\bf 74} 073102

\bibitem{P9}Hartmann H, Berggold K, Jodlauk S, Klassen I, Kordonis K, Fickenscher T,
P\"{o}ttgen R, Freimuth A and Lorenz T 2005 {\em J. Phys.: Condens.
Matter} {\bf 17} 7731

\bibitem{P10} Rohrkamp J 2007 {\em diploma thesis} Universit\"{a}t zu
K\"{o}ln

\bibitem{P11} Pierre J, de Combarieu A and Lagnier R 1979 {\em J. Phys. F: Metal Phys.} {\bf 9}
1271

\bibitem{P12} Chattopadhyay T, McIntyre G J, K\"{o}bler U 1996 {\em Solid State Com.} {\bf 100}
117

\end{thebibliography}
\end{document}